\documentclass[showpacs,twocolumn,pra]{revtex4}
\usepackage{amsmath}
\usepackage{amssymb}
\usepackage{xcolor}
\usepackage{graphicx}
\usepackage{soul}

\newcommand{\bra}[1]{\ensuremath{\left\langle #1 \right\vert}}
\newcommand{\ket}[1]{\ensuremath{\left\vert #1 \right\rangle}}

\newcommand{\boldvec}[1]{\ensuremath{\boldsymbol{\mathbf{#1}}}}
\newcommand{\spvec}[1]{\ensuremath{\mathbf{#1}}}
\newcommand{\unitvec}[1]{\ensuremath{\mathbf{\hat{#1}}}}
\newcommand{\sptensor}[1]{\ensuremath{\boldsymbol{\mathbf{#1}}}}

\newcommand{\eq}[1]{Eq.~\eqref{#1}}
\newcommand{\rv}{\spvec{r}}

\newcommand{\beq}{\begin{equation}}
\newcommand{\eeq}{\end{equation}}

\begin{document}
\title{Controlled manipulation of light by cooperative response of atoms in an optical lattice}
\author{Stewart D. Jenkins}
\affiliation{School of Mathematics, University of Southampton, Southampton SO17 1BJ,
  United Kingdom}
\author{Janne Ruostekoski}
\affiliation{School of Mathematics, University of Southampton, Southampton SO17 1BJ,
  United Kingdom}
\date{\today}
\begin{abstract}
  We show that a cooperative atom response in an optical lattice to
  resonant incident light can be employed for precise control and
  manipulation of light on a subwavelength scale.
  Specific collective excitation modes of the system that result from
  strong light-mediated dipole-dipole interactions  can be addressed
  by tailoring the spatial phase-profile of the incident light.
  We demonstrate how the collective response can be used to produce
  optical excitations at well-isolated sites on the lattice.
\end{abstract}
\pacs{03.75.Lm, 32.80.Qk}
\maketitle


Accurate control of ultracold atomic gases in periodic optical
lattices, in which interactions are well understood, opens the door to
unique and intriguing opportunities to study many-particle phenomena
and their applications.
Experimental progress has lead to observations of novel
strongly-interacting states, e.g., in quantum phase transitions
\cite{gremott,mott1d,tonks,JOR08,SCH08} and fermionic pair
condensation \cite{CHI06}.
Many-body quantum entanglement has been generated via controlled atom
collisions \cite{Mandel}, lattice systems have been used for
preparation of spin-squeezed states for sub-shot-noise interferometry
\cite{Esteve_Nature_2008}, and the atoms can now even be manipulated
in a single-spin level at a specific lattice site \cite{singlespin}.
On the other hand, recent developments in nanofabrication of arrays of
circuit elements acting as plasmonic resonators has stimulated
interest in photonic metamaterials.
A metamaterial is an artificially tailored crystal consisting of
subwavelength-scale structures that can manipulate light on a
nanoscale.
Here we show that a basic Mott-insulator state of a neutral gas of ultracold atoms confined in an
optical lattice, or artificial light crystal, exhibits strongly interacting electric dipole transitions
leading to a cooperative response. Such collective behavior can influence resonant imaging and
may also be employed to form a metamaterial for precise control and manipulation of optical fields on
a subwavelength scale, providing an interesting nanophotonic tool.

In this letter, we consider an ultracold gas of atoms confined in a
two-dimensional (2D)
optical lattice with precisely one atom per site.
Such a system can be prepared, e.g., in a weak harmonic trap or by engineering a
Mott-insulator state of atoms by single-site addressing of atomic
spins \cite{singlespin}.
Resonant, coherent light whose spatial phase-profile is adjusted, e.g., by a
hologram or spacial light modulator, illuminates the lattice.
The scattered light mediates strong many-particle
dipole-dipole (DD) interactions between atoms, leading to a cooperative
atom response.
The optical excitations of the atoms exhibit collective modes with
resonance frequencies and radiative linewidths that dramatically
differ from those of an isolated atom.
We demonstrate the idea of subwavelength-scale light manipulation by
engineering the spatial phase-profile of an incident monochromatic
plane-wave.
The tailored incident field produces localized dipolar
subwavelength-scale excitations of the atoms in desired locations in
the lattice.
By dynamically adjusting the phase pattern of the field, the
excitations can be controlled and moved around in the lattice.

The particular example of subwavelength-scale localization of optical
excitations we study here has attracted considerable interest in
nanophotonics with possibilities for microscopy and data storage applications.
Spatial and temporal modulation of ultrashort laser pulses leads 
to excitation of energy hot-spots in nanostructures
\cite{PhysRevLett.88.067402,Aeschlimann:ProceedingsOfTheNationalAcademyOfSciences:2010}.
It has also been proposed that interactions between induced currents and plasmonic waves on nanostructures permits the excitation of subwavelength hot-spots by amplitude or phase modulated
monochromatic fields \cite{PhysRevLett.101.013901,PhysRevLett.106.085501}.

Non-trivial collective optical properties result from a cooperative response
of the strongly interacting, closely-spaced atoms: recurrent
scattering events, in which a photon is repeatedly scattered by the
same atom, lead to collective modes with strongly modified spatial
configurations and radiation rates \cite{Lagendijk1996143,MoriceEtAlPRA1995,PhysRevA.55.513, PhysRevA.56.2056}.
Such scattering processes can result in 
light localization that is analogous
to Anderson localization of electrons 
\cite{vtiggelen99}.
The resonant response is very different from the studies of off-resonant optical diagnostics of atomic
correlations in optical lattices \cite{JAV03,MEK07,ruo_light09,rist_light,hulet_light,bur_light2,burhall}.
Photonic band gaps for atomic lattices have previously been calculated in Ref.~\cite{castinlattice}.

For our lattice system we numerically calculate the optical response
by stochastically sampling the atomic positions according to their
spatial distributions then solving the recurrent scattering events to
all orders for each stochastic realization.
We find a strong resonant response in the case of closely-spaced atoms
with the near-field emission pattern from the atoms forming sharp,
narrow spatially localized amplitude peaks.
The response is sensitive to detuning of the incident light from the
atomic resonance and to the spatial separation between the atoms.
Tuning light off-resonant or increasing the lattice spacing rapidly
leads to suppressed interactions.
If the atoms are not confined strongly enough to the individual
lattice sites, the resulting increased disorder in the atomic
positions due to quantum fluctuations also suppresses the strong collective effects in the
ensemble-averaged response.


We take the atoms to occupy the lowest energy band in a 2D square
optical lattice of periodicity $a$ in the $xy$ plane as described in
Appendix~\ref{sec:optic-latt-trapp}.
We assume that the atoms are tightly confined to the lowest
vibrational state in the $z$ direction of an oblate external potential
and that they reside in the Mott-insulator state with precisely
one atom per site.
In a combined harmonic trap and the lattice the single-occupancy state of bosonic atoms can
exist in a weak harmonic trap or can be
engineered from the typical `wedding-cake' Mott-insulator ground state
by manipulating the multi-occupancy states, e.g., by
single-site addressing or by inducing atom parity-dependent losses \cite{singlespin}.
In a deep lattice, the vibrational ground-state wavefunction (the Wannier
function) is approximately that of a harmonic oscillator with
frequency $\omega = 2\sqrt{s}E_R/\hbar$
where $s$ denotes the lattice depth in the units of the lattice-photon recoil
energy $E_R = \pi^2\hbar^2/(2ma^2)$ \cite{Morsch06}.
The confinement of the atoms along the lattice in the $xy$ plane in
each site $j$ thus has a Gaussian density profile, $\rho_j(\rv)$, with
the $1/e$ width $\ell = as^{-1/4} / \pi$ controlled by the
lattice depth $s$ and the lattice spacing $a$.

We illuminate the lattice with a monochromatic incident field
$\spvec{E}_{\mathrm{in}}(\rv,t)$ whose frequency $\Omega$ is nearly
resonant on an electric dipole transition.
This impinging field excites the dipole transition of the atoms,
producing scattered light that, in turn, impact the driving of
neighboring atoms and alter their scattered light.
The scattered photons can mediate strong interactions between closely-spaced atoms,
so that the atomic system responds to light cooperatively,
exhibiting {\em collective} excitation eigenmodes. 
Here, we show how to exploit these interactions for controlling and
manipulating light on a subwavelength scale.
As a specific example, we prepare 
subwavelength-scale spatially
localized collective excitations of the atoms in isolated regions of the lattice
by considering an incident plane-wave illumination of the atoms with
an approximately sinusoidal phase-profile.
Such a response is distinct from
that which would be seen if the atoms did not interact.

In order to model the cooperative atom response to light, we assume the
incident field is sufficiently weak that saturation of the excited
state can be neglected.
For simplicity, we consider the atomic internal states as an effective
two-level system consisting of a single electronic ground and excited
state.
On impact, light drives atomic transitions inducing a polarization
density $\spvec{P}^+(\rv,t) =\sum_j \spvec{P}_j^+(\rv,t) $, where the
polarization within each site $j$,  $\spvec{P}_j^+(\rv,t) =
\spvec{d}_{j}\delta(\rv - \rv_j)$ and $\spvec{d}_j$ is the electric
dipole moment of an atom at site $j$.
To facilitate numerical evaluation of the lattice response, we express
the polarization in terms of stochastic amplitudes $e_j$, representing
the coherence of atoms $j$ realized for a stochastic sampling of atomic
positions from the atomic density distributions $\rho_j(\rv_j)$, such that
$\spvec{P}_j^+(\rv,t) = e^{-i\Omega t} \wp \unitvec{d} \,\delta(\rv
- \rv_j) e_j(t)$, where $\wp$ is the atomic dipole matrix element.
When the atomic dynamics evolve on timescales much longer than the
light propagation time across the optical lattice
\cite{PhysRevA.55.513}, the induced polarization produces the
scattered electric displacement field
$\spvec{D}_{\mathrm{S},j}^+(\rv)= k^3/(4\pi) \int d^3r' \mathbf{G}(\rv
- \rv')\cdot \spvec{P}_{j}^+(\rv')$, where
$\mathbf{G}(\rv - \rv')$ is the monochromatic dipole radiation kernel representing the
radiated field at $\rv$ from a dipole residing at $\rv'$ \cite{Jackson}.
Thus, the atom at site $j$ experiences driving by the sum of fields
$\spvec{D}_{\mathrm{S,}j'}$ scattered from all other atoms in the
lattice and the incident field $\spvec{D}_{\mathrm{in}}$; these
scattered fields are proportional to the amplitudes $e_{j'}$ of their
atoms of origin.
These multiple scattering processes therefore
produce collective dynamics described by
\begin{align}
  \label{eq:excitedStateDynamicsMatrix}
  \dot{e}_j &= (i\delta - \Gamma/2)e_j + \sum_{j' \ne j}
  \mathcal{C}_{j,j'} e_{j'}   + F_j \textrm{, } \\
  \mathcal{C}_{j,j'} & \equiv
  \frac{3\Gamma}{2i}  \unitvec{d} \cdot
  \sptensor{G}(\rv_j - \rv_{j'}) \cdot \unitvec{d} \textrm{,}
\end{align}
where $F_j \equiv e^{i\Omega t}\wp
\unitvec{d} \cdot
\spvec{D}^+_{\mathrm{in}}(\rv_j)/(i\hbar\epsilon_0)$ is the
direct driving of the atom in site $j$ by the incident field, $\delta \equiv
\Omega - \omega_{e,g}$ is the detuning of the field from resonance,
and $\Gamma$ is the spontaneous emission rate.
For each stochastic realization of atomic positions, interactions
between $N$ atoms in an optical lattice lead to the formation of $N$
\emph{collective} atomic excitations, each with its own resonance
frequency and spontaneous emission rate, which could have either
superradiant or subradiant characteristics.
\begin{figure*}
  \centering
  \includegraphics{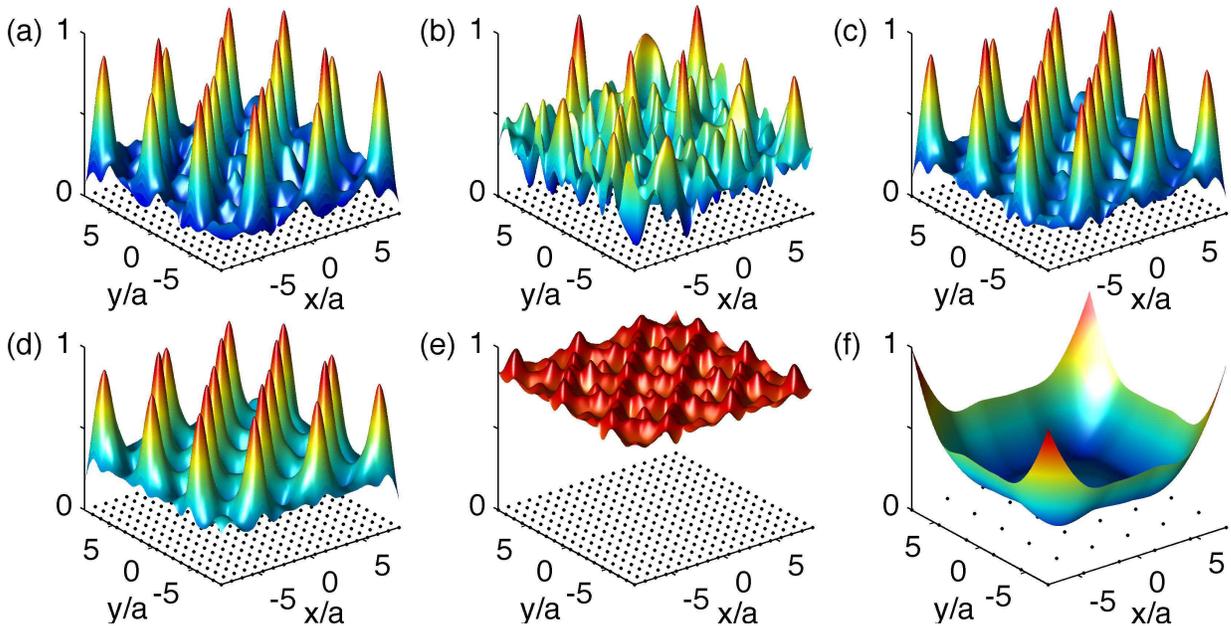}
  \caption{The atomic excitation intensities
    $\overline{|e_j|^2} / \max_{j'} \overline{|e_{j'}|^2}\rangle$
    resulting from the cooperative response of an
    ensemble of two-level atoms in an optical lattice to a phase
    modulated incident field [Eq.~\eqref{eq:pmElecfield}]
    with modulation period of $6$ lattice sites.
    The black dots indicate the positions of the populated lattice
    sites.
    In panel (a), the atoms are perfectly confined at the center of
    each site and their wavefunctions have width $\ell=0$.
    Panel (b) represents the response of a single stochastic
    realization of atomic positions sampled from a 2D
    Gaussian variables of width $0.12\,a$ centered on the lattice sites.
    Panels (c), (e-f) the show responses of atoms whose wavefunction widths
    $\ell = 0.12\,a$ calculated from an ensemble average over several thousand
    realizations of atomic positions, while in panel (d) $\ell = 0.21\, a$.
    In panel (e), the incident field is detuned from the
    resonance of an isolated atom by $10\Gamma$, while the detuning is zero for all other
    panels.
    In panel (f), atoms were removed from 8/9 of the sites, resulting in
    an effective spacing of $a'=3a$, thus removing the role of
    the other sites from the cooperative response and destroying the
    pattern observed in panels (a) and (c).
  }
  \label{fig:Response}
\end{figure*}

By exploiting the strong DD interactions, one can tailor
the incident field so that it excites specific linear combinations of
collective modes, providing a desired response.
Here, for example, we consider a linearly-polarized ($\unitvec{e}_y$ chosen to be parallel to the lattice plane), phase-modulated
field whose positive frequency component reads
\begin{subequations}
  \label{eq:pmElecfield}
  \begin{align}
   \spvec{E}_{\mathrm{in}}^{+}(\rv,t) \simeq \unitvec{e}_y E_0
    e^{i\varphi(x,y)} e^{i(kz-\Omega t)} \textrm{,} \\
    \varphi(x,y) = {\pi\over 2} \sin(\kappa x) \sin(\kappa
    y) \,.
  \end{align}
\end{subequations}
A field profile of this form, however, contains evanescent plane-wave
components whose transverse wavevectors exceed the carrier wave number
$k=2\pi/\lambda=\Omega/c$.
We therefore approximate it through the truncated Fourier expansion as
described in Appendix~\ref{sec:incid-electr-field}.
One can produce such a field, e.g., by 
a hologram or a
spatial light modulator. 
Figure~\ref{fig:Response} illustrates how the 
field
\eq{eq:pmElecfield} can excite a checkerboard pattern of localized
excitations of atoms in 
a lattice.
In these calculations, we neglect the width of the atomic
wavefunctions in the $z$ direction, as a width on the order of
$\ell$ in the $z$ direction has a negligible effect on the calculated
response.
We fully incorporate, however, quantum fluctuations of the atomic positions
on the lattice plane in the vibrational ground state of each site.
The many-atom correlations of the specific one-atom Mott-state are also
included.
As we discuss below,
interactions between the atoms 
are vital to the realization of this pattern.
The atoms are arranged in an $18 \times 18$ square lattice with the spacing
$a =  0.55\lambda$, and a site residing at $(x_0,y_0) =(a/2,a/2)$.
The incident field [Eq.~\eqref{eq:pmElecfield}] has a modulation period $2\pi/\kappa =
6a$, indicating the periodicity of $\spvec{E}_{\mathrm{in}}(\rv,t)$ of
six sites.
We choose the dipole orientation $\unitvec{d} \approx \unitvec{e}_z + 0.1
\unitvec{e}_y$ to be slightly rotated from the normal to the lattice
plane so that the atoms scatter fields largely within the plane while
allowing them to be driven by the incident field.
The collective mode spontaneous emission rates  
range from a very subradiant $3\times 10^{-3}\Gamma$
to the superradiant $5\Gamma$, while their frequency shifts from the
single atom resonance range from $-2\Gamma$ to $0.9\Gamma$.

We first consider an infinitely deep lattice in which the atoms
are perfectly confined at the center in their respective sites,
i.e., with $\ell=0$.
In this case, the atomic positions are deterministic and
Eq.~\eqref{eq:excitedStateDynamicsMatrix} reduces to a coupled set of
linear equations whose steady-state solutions for $|e_j|^2$ are shown in
Fig.~\ref{fig:Response}a.
A subset of atoms residing at the local minima of $\varphi(x,y)$ are
more strongly excited than those in the surrounding lattice sites,
while those at the local maxima of $\varphi(x,y)$ are roughly as
weakly excited as their surroundings.
The peaks sit on a background with excitations roughly $0.2$ times
those of the most excited atoms.
We find a subwavelength excitation FWHM width of the peak to be less than
$0.9\lambda$.
This results in a checkerboard pattern with the localized excitations
separated by $3\sqrt{2}$ sites ($3$ sites in the both directions)
sitting on a background of weakly excited atoms.
The regularity of the response can be broken by slightly altering the
period or the orientation of the phase modulation.
Note that the periodicity of the incident
field is significantly larger than the width of the localized
excitations.

In a more realistic scenario, the width of the lattice site atomic wavefunction due to vacuum fluctuations
cannot be neglected and the scattered light sources are essentially distributed over the atomic
densities.
To obtain the average atomic excitations $\overline{|e|_j^2}$ that
dominate the near-field emission, we solve
Eq.~\eqref{eq:excitedStateDynamicsMatrix} through Monte-Carlo
integration \cite{JavanainenEtAlPRA1999}.
We obtain a large number of realizations of atomic positions  $\rv_j$ 
in each site $j$   
sampled from a probability distribution matching the single-atom
density function.
Then for each realization, we solve
Eq.~\eqref{eq:excitedStateDynamicsMatrix} as if the atoms were
localized at the sample points.
We then compute $|e_j|^2$ for each sample, and perform an ensemble
average over all realizations of atomic positions.
Each stochastic realization represents a possible outcome of a single
experimental run in which case atomic positions are localized due to
the measurement of scattered photons.

Quantum 
fluctuations of atomic positions in individual sites can dramatically affect
the response.
We demonstrate this showing in Fig.~\ref{fig:Response}b the excitation
intensities for a single stochastic realization of atomic positions in
which case a single atom in each site is independently sampled from
the Gaussian density-distribution of width $\ell = 0.12\,a$
(corresponding to the lattice height $s\simeq50$).
The non-regularity of the lattice alters the collective modes for the
sample realization.
Although the atoms are driven by the same incident field that gives
rise to the pattern in Fig.~\ref{fig:Response}a, the displacement
drastically alters the collective interaction, and yields an optical response
with a significant stochastic noise and a less regular array of localized peaks.
Such effects can be washed out when one calculates the
ensemble average of the response that corresponds to expectation values
obtained over many experimental realizations.
The excitation intensity $\overline{|e_j|^2}$ averaged over 6400
position realizations  for Wannier functions of width $\ell = 0.12\,a$
is shown in Fig.~\ref{fig:Response}c for the same parameters as those
in Fig.~\ref{fig:Response}a.
With 
atomic wavefunction of this width, the collective
interactions 
producing the pattern of Fig.~\ref{fig:Response}a survive
the averaging process, providing an excitation 
with subwavelength
FWHM 
and a background excitation comparable to that calculated
for perfectly localized atoms in an infinitely deep lattice.
Weaker confinement only moderately diminishes the cooperative
interactions.
For $\ell = 0.21 a$, corresponding to $s\simeq 5$,
Fig.~\ref{fig:Response}d shows a weakening contrast of the
pattern with a background excitation approximately $0.3$ 
times that of
the peaks, which themselves have slightly broader FWHM of $1.2\lambda$.

We can illustrate the essential nature of cooperative interactions in
the formation of this excitation pattern by suppressing the DD
interactions and observing the result.
Fig.~\ref{fig:Response}e shows the response of the atoms in the
optical lattice of Fig.~\ref{fig:Response}c to an incident
field detuned from the single atom resonance by $10\Gamma$.
Since, with this detuning, each photon interacts more weakly with an
individual atom, the probability of a multiple scattering process for a two-level atom
reduces geometrically with the number of single scattering events
involved in that process \cite{PhysRevA.56.2056}.
The 
interactions resulting from
these multiple scattering events are therefore suppressed.
While the incident field has the same phase modulation as in
Figs.~\ref{fig:Response}a-d, the excitation pattern is not preserved.
Atoms at sites where the phase modulation $\varphi$ is minimized
are not appreciably more excited than the surrounding background as in
Figs. \ref{fig:Response}a and c.

The pattern formation in Figs.~\ref{fig:Response}a,c truly represents
a cooperative response where the interactions between all the atoms in
the lattice, including also the weakly excited ones, are essential.
We illustrate this in Fig.~\ref{fig:Response}f, where all of the atoms
not residing at the maxima or minima of the incident field phase
modulation $\varphi$ have been removed from the lattice.
In effect, the lattice spacing was tripled to $a'=3a = 1.65\lambda$,
with an atom residing at $(x_0,y_0) = (a'/2, a'/2)$.
If the 
removed 
atoms
had not played an essential
role, the response of the lattice would show 
a checkerboard pattern
of strongly excited and weakly excited atoms.
However, the system response displayed in Fig.~\ref{fig:Response}f
shows
that 
the atoms 
in the interior of the lattice
are excited 
roughly evenly even though
each atom is driven with an opposite phase to that of its nearest
neighbor. 
Atoms at the edge of the sample are more strongly excited due to finite size effects.

Sharp localized excitations may be broadened by heating and losses
that can inhibit the cooperative atom response.
Raman transitions to other vibrational center-of-mass states heat up the atoms,
broadening the atomic density distributions in individual sites and
increasing the hopping amplitude of the atoms between the adjacent
sites.
Such processes could be reduced, e.g., due to the orthogonality of
the eigenfunctions in each site, if the electronic ground and excited
state atoms approximately experience the same lattice potential even if the system
is not in the Lamb-Dicke regime.
Alternatively, if the atomic linewidth is much larger than the
trapping frequency, the collective response may reach a steady state
before the heating becomes deleterious.
Additional atom losses due to Raman transitions to other internal states
could be avoided if each atom forms an effective two-level system of a
single electronic ground and excited state.
A desired two-level configuration could be realized with a cycling
transition by shifting all other transitions out of resonance or with a
$J=0\rightarrow J=1$ transition with all except one excited state shifted
out of resonance.

In conclusion, we have shown that resonant DD interactions
between atoms in an optical lattice lead to a collective
response that can be exploited in manipulation of light
on a subwavelength scale.
To illustrate this, we 
studied an example of engineering a checkerboard pattern of isolated atomic excitations.
Unlike in nanofabricated metamaterial samples \cite{PhysRevLett.88.067402,Aeschlimann:ProceedingsOfTheNationalAcademyOfSciences:2010,PhysRevLett.101.013901,PhysRevLett.106.085501},
here the effect is not based on interactions between 
plasmonic and current excitations
 but purely electric DD interactions between neutral atoms without magnetic contributions.
Moreover, the positions of the excitations can be dynamically altered simply by
translating the phase-modulation pattern, so that the collective excitation pattern
adiabatically follows the change in the phase pattern.
By understanding these interactions, the characteristics of an
incident field could be engineered to produce more complex
excitations. 
Our example also demonstrates how a cooperative response can have 
implications on the resonant
absorption imaging of 2D atomic samples in which case deviations from the column
density results have been experimentally observed \cite{Chomaz11}. Moreover, the
non-trivial relationship between incident field modes
and the collective excitations to which they couple could be of
importance, e.g., to imaging and to the implementation of quantum memories in optical
lattice systems in which the states of a light field are mapped onto
collective hyperfine excitations in the lattice.

\acknowledgments{We acknowledge financial support from the Leverhulme
  Trust and the EPSRC.}

\appendix

\section*{Appendix}

In these appendices, we provide some of the technical
details of the optical lattice system and the incident field used to
excite it.
In Appendix~\ref{sec:optic-latt-trapp}, we describe the optical lattice
potential.
In Appendix~\ref{sec:dynam-atom-excit}, we describe the
dynamics of the atomic dipoles interacting with both the incident and
scattered electric fields.
Finally in Appendix~\ref{sec:incid-electr-field}, we describe how one
can construct the approximate phase modulated driving field used to
illuminate the optical lattice system and excite collective atomic
excitations that were discussed in the text.

\section{The Optical Lattice Trapping Potential}
\label{sec:optic-latt-trapp}

We consider a two-dimensional square optical lattice of
periodicity $a$, centered on the point $(x_0,y_0,0)$.
Four intersecting beams produce the optical confining potential
\begin{equation}
  \label{eq:V_lat}
  V = sE_R\left[\sin^2\left(\pi \frac{x-x_0}{a}\right) +
    \sin^2\left(\pi \frac{y-y_0}{a}\right) \right]
\end{equation}
where $E_R = \pi^2\hbar^2/(2ma^2)$ is the lattice recoil energy \cite{Morsch06}, and
the dimensionless confinement strength $s$ controls the width of the vibrational ground-state
wavefunction in each site (Wannier functions) that results from single-particle quantum fluctuations.
An additional potential tightly confines the atoms in the $z=0$ plane.
The lattice resides in a Mott-insulator state with precisely
one atom per lattice site.
Each lattice site, labeled by index $j$, has a potential minimum
located at position $\spvec{R}_j$, and the Wannier function
$\phi_j(\rv) \equiv \phi(\rv - \spvec{R}_j)$.
When the confinement is sufficiently tight, the potential $V$ in the
neighborhood of each lattice site is roughly harmonic with
\begin{equation}
  \label{eq:V_lat_approx}
  V(\rv) \approx  \frac{m}{2}\sum_{\mu=x,y,z} \omega_\mu^2 (\Delta r_\mu)^2
\end{equation}
where $\omega_x = \omega_y = 2\sqrt{s}E_R/\hbar$, $\omega_z$ is the
trapping frequency along the $z$ direction, and $\Delta r_\mu
\equiv \unitvec{e}_\mu \cdot (\rv - \spvec{R}_j)$ is the displacement of the
$\mu$th component of $\rv$ from the lattice site $\spvec{R}_j$.
An atom in each site occupies the ground state of the harmonic
oscillator potential
\begin{equation}
  \label{eq:phi}
  \phi(\rv) = \frac{1}{(\pi^3\ell^4\ell_z^2)^{1/4}}
  \exp\left(-\frac{x^2+y^2}{2\ell^2} - \frac{z^2}{2\ell_z^2}\right) \text{,}
\end{equation}
where the width
of the of the wave function is $\ell = as^{-1/4} / \pi$, and
its thickness $\ell_z = \sqrt{\hbar/(m\omega_z)}$.
The atomic density, $\rho_j(\rv) \equiv |\phi_j(\rv)|^2$, at site
$j$  thus has a Gaussian profile with a $1/e$ radius
$\ell$ in the $xy$ plane.
This width is directly proportional to the lattice spacing and narrows
with increased trapping strength $s$. It is possible to adjust the lattice spacing
by using accordion lattices \cite{AlAssam10}. Moreover, the relationship between the lattice spacing
and the Wannier wavefunction confinement may be controlled by magnetic field dressing \cite{Shotter08}.

\section{Dynamics of atomic excitations in the lattice}
\label{sec:dynam-atom-excit}

In this section, we elaborate on the dynamics of the atomic electric
dipoles interacting with light.
We show that
the evolution of the system can be described as in Eq.~(2) of the main
text.
We then discuss the Monte-Carlo method used to compute
the average excitation energies $\overline{|e_j|^2}$ displayed in
Fig. 1 of the text.

\subsection{Basic model}

We consider an ensemble $N$ two-level atoms placed in harmonic
potentials centered at the lattice sites $\spvec{R}_j$ ($j=1,\ldots,N$).
Each atom has an internal ground state $\ket{g}$ and an excited state
$\ket{e}$ which differ in energy by $\hbar\omega_{e,g}$.
A coherent,  monochromatic field
$\boldvec{D}^+_{\mathrm{in}}(\rv,t)$ with frequency $\Omega$
impinges on the lattice.
The field incident on the atoms in the lattice drives the $\ket{g}
\leftrightarrow \ket{e}$ transition, inducing a polarization density
$\spvec{P}^+(\rv,t) = \sum_j \spvec{P}_j^+(\rv,t)$, where the
polarization due to an atom in site $j$, $\spvec{P}_j^+ = \spvec{d}_j
\delta(\rv - \rv_j)$, $\spvec{d}_j$ is the dipole moment of an atom
in site $j$, and $\rv_j$ is its position coordinate.
Scattering of light from the atoms produces
correlations between the positions of the atoms and the scattered
field profile.
In this way, continuous measurement of the scattered light yields
an effective measurement of the atomic positions.
This process simultaneously projects the position coordinates of an atom in each site $j$ onto a
specific positon $\spvec{r}_j$ through the projection operator
$\hat{Q}(\{\rv_{j}\} )  \equiv \bigotimes_{j=1}^N
\ket{\rv_{j}}_{j}\bra{\rv_{j}}$ acting on the initial vibrational
state of the system, where $\{\rv_j\}$ denotes the set
$\{\rv_1,\ldots,\rv_N\}$ of realized position coordinates for atoms in
lattice sites $\{1,\ldots,N\}$.
We therefore characterize the evolution of the polarization
density associated with site $j$ in terms of
the slowly varying coherence conditioned on the realized atomic
position $\rv_j$, and the positions in all other sites $\rv_{j'}$ for  $j' \ne j$.
We define the amplitudes $e_j(\rv_j,t;\{\rv_{j'\ne j}\}) \equiv
e^{i\Omega t} \langle \sigma_j \rangle_{\{\rv_{1},\rv_2,\ldots,\rv_{N}\}}$ to be the
slowly varying coherence between the ground and excited states of an
atom at site $j$ conditioned on the observation of atoms at positions
$\{\rv_{1},\rv_{2},\ldots,\rv_{N}\}$ within their respective sites,
where $\sigma_j$ is the  coherence operator for
an atom in site $j$.
For a single realization of position coordinates of the atoms, the
polarization in site $j$ depends on the position of all the atoms via
the amplitudes $e_j$:
\begin{equation}
  \label{eq:Pol_of_at_j}
  \boldvec{P}_j(\rv,t; \{\rv_{j'}\}) =  e^{-i\Omega t}\unitvec{d} \wp \delta(\rv - \rv_j)
  e_j(\rv_j,t; \{\rv_{j'\ne j} \}) \text{,}
\end{equation}
where $\wp$ is the dipole matrix element.
In addition to the incident electric field, an atom at site $j$ experiences
driving by the fields $\boldvec{D}^+_{\mathrm{S,}j'}$ scattered from all
other sites $j'\ne j$ in the lattice.
The total displacement field arriving at this site is thus
\begin{equation}
  \boldvec{{D}}^+_{\mathrm{ext},j}(\rv,t) \equiv
  \boldvec{D}^+_{\mathrm{in}}(\rv,t) + \sum_{j'\ne j}
  \boldvec{D}^+_{\mathrm{S},j'}(\rv,t).\label{eq:Dext}
\end{equation}
When the atomic dynamics evolve on timescales much longer than the
time it takes for light to propagate across the optical lattice, the
scattered fields can be expressed in the monochromatic limit as
\cite{PhysRevA.55.513, PhysRevA.56.2056}
\begin{equation}
  \label{eq:5}
  \boldvec{D}^+_{\mathrm{S},j}(\rv)= \frac{k^3}{4\pi} \int d^3r'
  \mathbf{G}(\rv - \rv')\cdot \boldvec{P}^+_{j}(\rv') \text{,}
\end{equation}
where
$k=\Omega/c$, and $\mathbf{G}$ is the radiation kernel with tensor
components
\begin{equation}
  \mathbf{G}_{\mu,\nu}(\rv) = \frac{1}{k^2}
  \left(\partial_\mu\partial_\nu - \delta_{\mu,\nu}
    \nabla^2\right)\frac{e^{ikr}}{kr}\,.
  \label{eq:RadKernel}
\end{equation}
The monochromatic dipole radiation kernel, representing the radiated field at ${\bf r}$ from
a dipole residing at ${\bf r}'$, can be expressed as
\cite{Jackson}
\begin{align}
  \label{eq:RadKer2}
  \sptensor{G}(\rv) = & \left[\frac{2}{3} +
    \left(\frac{\rv\rv}{r^2} - \frac{1}{3}\right)
    \left(\frac{3-3ikr-k^2r^2}{k^2 r^2} \right)\right]
  \frac{e^{ikr}}{kr}  \nonumber \\
  & + \frac{8\pi}{3}\delta(k\rv)\text{.}
\end{align}

\subsection{Stochastic simulations}

In the simulations we solve the cooperative optical response of the
atomic sample to the incident field using a Monte-Carlo approach in
which the position coordinates of the atoms are sampled according to their
position correlation functions, and the optical response is calculated
for each stochastic realization \cite{JavanainenEtAlPRA1999}.
We assume that the atoms form a Mott-insulator state of precisely one
atom per lattice site.
The position coordinates of atoms within each site $j$ are therefore independent stochastic
variables, sampled from Gaussian distributions $\rho_j(\rv)$. The atoms reside at the vibrational
ground states of the lattice sites and the Gaussian distribution $\rho_j(\rv)$ results from the
single-particle quantum fluctuations, determined by the Wannier wavefunction
density in the particular site $j$. The atomic position in the site $j$
is centered at $\spvec{R}_j$. The width of the Wannier wavefunction is determined by the lattice confinement.
Since each stochastic realization of position coordinates for the atoms
in the $N$ lattice sites $\{\rv_1,\rv_2, \ldots, \rv_N\}$ can be
interpreted as an outcome of a continuous measurement process of
scattered light that localizes the atoms, each stochastic Monte-Carlo
trajectory also represents a possible outcome of a single experimental
run.
We evaluate the ensemble averages of the atomic optical excitations
by computing the excitations for each of many stochastic realizations
of position coordinates of the atoms and then calculating their ensemble average.
The numerically calculated ensemble average then corresponds to the
experimentally measured ensemble average over many experimental runs
and provides quantum mechanical expectation values of the
observables.

In order to calculate the optical response for each stochastic
realization of position coordinates for the atoms in the $N$ lattice sites,
$\{\rv_1,\rv_2, \ldots, \rv_N\}$, we assume here that the incident
light is sufficiently weak that we can neglect the excited state
saturation of the atoms.
In this limit, the coherence amplitude for an atom in site $j$ obeys the
equations of motion
\begin{align}
  \frac{d}{dt} e_j = (i\delta -
  \Gamma/2)e_j  + e^{i\Omega t}
  \frac{\wp}{i\hbar\epsilon_0}
  \unitvec{d} \cdot \boldvec{D}^+_{\mathrm{ext},j}(\rv_j,t)
  \textrm{, }
  \label{eq:excitedStateDynamics}
\end{align}
where $\delta \equiv \Omega - \omega_{e,g}$ is the detuning of the
incident light from the resonance of the atomic transition and
$\Gamma$ is the atomic spontaneous emission rate.
For each stochastic realization of atomic positions, the light
impinging on an atom in a particular site consists of the incident
field and the scattered light from all the other $N-1$ sites.
Because light scattered from an atom is directly proportional to its coherence
amplitude, we obtain
\begin{equation}
  \frac{d}{dt} e_j = (i\delta - \Gamma/2)e_j
  + \sum_{j' \ne j}
  \mathcal{C}_{j,j'} e_{j'}  + e^{i\Omega t} \frac{\wp}{i\hbar\epsilon_0}
  \unitvec{d} \cdot \boldvec{D}^+_{\mathrm{in}}(\rv_j,t)
  \textrm{, }
  \label{eq:excitedStateDynamicsMatrixApp}
\end{equation}
where $\mathcal{C}_{j,j'}$ is the coupling matrix element representing
the effect of dipole radiation from the atom in site $j'$ on the
coherence amplitude of the atom in site $j$;
\begin{equation}
  \label{eq:MatrixElements}
  \mathcal{C}_{j,j'} =
  \frac{3\Gamma}{2i} \unitvec{d} \cdot \sptensor{G}(\rv_j - \rv_{j'})
  \cdot \unitvec{d} \text{.}
\end{equation}
For each stochastic realization of position coordinates of the atoms,
\eq{eq:excitedStateDynamicsMatrixApp} represents a collective response of
the atomic sample to the incident light.
Repeated exchanges of a photon between the same atoms lead to
cooperative response of the atoms.
The system exhibits collective eigenmodes with each eigenmode
associated with a specific resonance frequency and a radiative
linewidth.
In the text, we considered an $18\times 18$ array with lattice spacing
$0.55\lambda$ ($\lambda = 2\pi c/\Omega$), and dipole orientations
$\unitvec{d} \approx \unitvec{e}_z + 0.1 \unitvec{e}_y$.
For this configuration, when the atoms are placed precisely at the
centers of the lattice sites $\spvec{R}_j$, the collective mode
spontaneous emission rates range from a very subradiant $3\times
10^{-3}\Gamma$
to the superradiant $5\Gamma$, while their collective frequency shifts
from the single atom resonance range from $-2\Gamma$ to $0.9\Gamma$.

In each stochastic realization of position coordinates of atoms within
their respective lattice sites, we solve the
steady-state solution of the excitation amplitudes of the atom in each
lattice site $e_j$ by setting $d e_j/dt =0 $ in
\eq{eq:excitedStateDynamicsMatrixApp}.
The results for a single stochastic realization of $|e_j|^2$ are
displayed in the main text.
We then compute the ensemble average of the atomic excitation
$\overline{|e_j|^2}$ by averaging over a sufficiently large number of
stochastic realizations.
That is, we take $\mathcal{N}$ stochastic realizations of the position
coordinate of an atom within each site $j$.
For the $\nu$th realization ($\nu = 1,\ldots,\mathcal{N}$) we have
sampled the position coordinates of an atom in site $j$,  $\rv_j^{(\nu)}$.
For these atomic positions, we solve the steady state of
Eq.~\eqref{eq:excitedStateDynamicsMatrixApp} to obtain the
atomic amplitudes $e_j^{(\nu)}$ and calculate the ensemble average
\begin{equation}
  \label{eq:1}
  \overline{|e_j|^2} = \frac{1}{\mathcal{N}}
  \sum_{\nu=1}^{\mathcal{N}} \left|e_j^{(\nu)}\left(\rv_j^{(\nu)}; \left\{\rv_{j'\ne
    j}^{(\nu)}\right\}\right)\right|^2\,.
\end{equation}


\subsection{Many-particle correlations}

In this paper we consider a specific many-particle atom state: the
bosonic Mott-insulator state with precisely one atom per site. The atoms
reside in the lowest vibrational state of each lattice site. Vacuum fluctuations
for the position of the single-particle state of the atoms are incorporated in the Monte-Carlo
sampling of the atomic positions from the Gaussian density distribution, determined by
the Wannier wavefunctions of each lattice site. As we will show below, for the specific
Mott-state of one atom per site, the sampling procedure also represents all the many-particle
correlations of the system.

The joint probability distribution $P(\bar\rv_1,\ldots,\bar\rv_N)$
for the positions of the $N$ atoms is given by
the absolute square of the normalized many-particle wavefunction
\begin{align}
P(\bar\rv_1,\ldots,\bar\rv_N) &= |\Psi (\bar\rv_1,\ldots,\bar\rv_N)|^2\\
P(\bar\rv_1,\ldots,\bar\rv_N) &\simeq {1\over N!}\sum_{j_1\ldots j_N} |\phi_{j_1} (\bar\rv_1) |^2\ldots |\phi_{j_N} (\bar\rv_N) |^2 \label{manybody}
\end{align}
Here the coordinates of the $N$ atoms are denoted by $( \bar \rv_1, \ldots,\bar \rv_N)$. The summation runs over all possible $N$-tuples $(j_1,\ldots,j_N)$ of the state labels $j=(1,\ldots,N)$ referring to the lattice sites $j$ with precisely one atom per site. The Wannier wavefunctions for each lattice site $\phi_j$ in Eq.~(\ref{manybody}) are assumed to have a negligible overlap with the neighboring sites, so that $\phi_j(\rv)\phi_k(\rv)\simeq 0$, whenever $j\neq k$. Since we have exactly one atom per site, we may rewrite the joint probability distribution of Eq.~(\ref{manybody}) as
\beq
P(\bar\rv_1,\ldots,\bar\rv_N) \simeq {1\over N!}\sum_{i_1\ldots i_N} |\phi_{1} (\bar\rv_{i_1}) |^2\ldots |\phi_{N} (\bar\rv_{i_N}) |^2 \label{manybody2}
\eeq
where the summation runs over all possible permutations of the atomic coordinates $i=(1,\ldots,N)$. Since there are $N!$ such permutations, we can express the joint probability distribution in terms of the positions of an atom in the $j$th lattice site $\rv_j$, instead of the position coordinates of the $j$th atom $\bar\rv_j$. Each $N!$ term contributes equally in the sum and we obtain for the joint probability distribution of the position of an atom in the lattice sites $j=(1,\ldots,N)$ in the coordinate representation $(\rv_1,\ldots,\rv_N)$
\beq
P'(\rv_1,\ldots,\rv_N) \simeq |\phi_{1} (\rv_{1}) |^2\ldots |\phi_{N} (\rv_{N}) |^2 \label{manybody3}
\eeq
In the resulting joint probability distribution the position coordinates of atoms within each site $j$ are independent stochastic variables that are sampled from the Gaussian distributions $\rho_j(\rv) = |\phi_{j} (\rv) |^2$, corresponding to the stochastic Monte-Carlo sampling procedure implemented in the previous section.

\section{The incident light}
\label{sec:incid-electr-field}

By exploiting interactions between the atoms, one can tailor the
incident field so that it drives specific linear combinations of
collective modes, providing a desired collective response.
In the text, for example, we consider phase modulated driving that can
excite an array of isolated atoms arranged in a checkerboard pattern
on an optical lattice.
A spatial light modulator is employed to produce an incident field
approximately of the form
\begin{equation}
  \label{eq:pmElecfieldApp}
  \spvec{{E}}_{\mathrm{in}}^{+}(\rv,t) = \unitvec{e}_y E_0
  e^{i(kz-\Omega t)}
  e^{i\varphi(x,y)} \textrm{,}
\end{equation}
where
\begin{equation}
  \label{eq:3}
  \varphi(x,y) = \varphi_{\mathrm{max}} \sin(\kappa x) \sin(\kappa y)
  \textrm{.}
\end{equation}
A field profile of this form, however, contains evanescent plane wave
components whose transverse wavevectors excede the the carrier
wave number $k$.
We therefore approximate this phase modulated field through the
truncated Fourier expansion
\begin{equation}
  \label{eq:incidentFieldTruncated}
  \spvec{E}_{\mathrm{in}}^{+} = \sum_{m,n} C_{m,n} \unitvec{e}_{m,n}
  e^{i(\spvec{k}_{m,n} \cdot \spvec{r} - \Omega t)}\textrm{,}
\end{equation}
where $\spvec{k}_{m,n} = m \kappa \unitvec{e}_x + n
\kappa \unitvec{e}_y + q_{m,n} \unitvec{e}_z$, $q_{m,n} \equiv
\sqrt{(\Omega/c)^2 - \kappa^2(m^2+n^2)}$,
and $\unitvec{e}_{m,n}$ 
is the normalized
projection of the vector $\unitvec{e}_y$ onto the plane perpendicular to
$\spvec{k}_{m,n}$.
We have truncated the expansion for values of $m$ and $n$ for which
$q_{m,n}^2\ge 0$.

Because our goal was to produce a phase modulated driving of the
meta-atoms, we choose $C_{m,n}$ so as to reproduce a phase modulated
driving, i.e.
\begin{align}
  \label{eq:4}
  \unitvec{d}\cdot \spvec{E}_{\mathrm{in}}^+ &= \sum_{m,n} C_{m,n}
  (\unitvec{d}\cdot\unitvec{e}_{m,n})
  e^{i(\spvec{k}_{m,n} \cdot \spvec{r} - \Omega t)}  \\
  &\approx \unitvec{d} \cdot \unitvec{e}_y e^{i\varphi(x,y)}
  e^{i(kz-\Omega t)}\textrm{,}
\end{align}
The coefficients $C_{m,n}$ are obtained from the discrete Fourier
transform of the phase modulation $\exp(i\varphi(x,y))$ and dividing
by $\unitvec{e}_{m,n}\cdot \unitvec{d}$.


\begin{thebibliography}{99}

\expandafter\ifx\csname natexlab\endcsname\relax\def\natexlab#1{#1}\fi
\expandafter\ifx\csname bibnamefont\endcsname\relax
  \def\bibnamefont#1{#1}\fi
\expandafter\ifx\csname bibfnamefont\endcsname\relax
  \def\bibfnamefont#1{#1}\fi
\expandafter\ifx\csname citenamefont\endcsname\relax
  \def\citenamefont#1{#1}\fi
\expandafter\ifx\csname url\endcsname\relax
  \def\url#1{\texttt{#1}}\fi
\expandafter\ifx\csname urlprefix\endcsname\relax\def\urlprefix{URL }\fi
\providecommand{\bibinfo}[2]{#2}
\providecommand{\eprint}[2][]{\url{#2}}

\bibitem[{\citenamefont{Greiner et~al.}(2002)\citenamefont{Greiner, Mandel,
  Esslinger, H\"{a}nsch, and Bloch}}]{gremott}
\bibinfo{author}{\bibfnamefont{M.}~\bibnamefont{Greiner}},
  \bibinfo{author}{\bibfnamefont{O.}~\bibnamefont{Mandel}},
  \bibinfo{author}{\bibfnamefont{T.}~\bibnamefont{Esslinger}},
  \bibinfo{author}{\bibfnamefont{T.~W.} \bibnamefont{H\"{a}nsch}},
  \bibnamefont{and} \bibinfo{author}{\bibfnamefont{I.}~\bibnamefont{Bloch}},
  \bibinfo{journal}{Nature} \textbf{\bibinfo{volume}{415}}, \bibinfo{pages}{39}
  (\bibinfo{year}{2002}).

\bibitem[{\citenamefont{St\"oferle et~al.}(2004)\citenamefont{St\"oferle,
  Moritz, Schori, K\"ohl, and Esslinger}}]{mott1d}
\bibinfo{author}{\bibfnamefont{T.}~\bibnamefont{St\"oferle}},
  \bibinfo{author}{\bibfnamefont{H.}~\bibnamefont{Moritz}},
  \bibinfo{author}{\bibfnamefont{C.}~\bibnamefont{Schori}},
  \bibinfo{author}{\bibfnamefont{M.}~\bibnamefont{K\"ohl}}, \bibnamefont{and}
  \bibinfo{author}{\bibfnamefont{T.}~\bibnamefont{Esslinger}},
  \bibinfo{journal}{Phys. Rev. Lett.} \textbf{\bibinfo{volume}{92}},
  \bibinfo{pages}{130403} (\bibinfo{year}{2004}).

\bibitem[{\citenamefont{Paredes et~al.}(2004)\citenamefont{Paredes, Widera,
  Murg, Mandel, F\"{o}lling, Cirac, Shlyapnikov, H\"{a}nsch, and
  Bloch}}]{tonks}
\bibinfo{author}{\bibfnamefont{B.}~\bibnamefont{Paredes}},
  \bibinfo{author}{\bibfnamefont{A.}~\bibnamefont{Widera}},
  \bibinfo{author}{\bibfnamefont{V.}~\bibnamefont{Murg}},
  \bibinfo{author}{\bibfnamefont{O.}~\bibnamefont{Mandel}},
  \bibinfo{author}{\bibfnamefont{S.}~\bibnamefont{F\"{o}lling}},
  \bibinfo{author}{\bibfnamefont{I.}~\bibnamefont{Cirac}},
  \bibinfo{author}{\bibfnamefont{G.~V.} \bibnamefont{Shlyapnikov}},
  \bibinfo{author}{\bibfnamefont{T.~W.} \bibnamefont{H\"{a}nsch}},
  \bibnamefont{and} \bibinfo{author}{\bibfnamefont{I.}~\bibnamefont{Bloch}},
  \bibinfo{journal}{Nature} \textbf{\bibinfo{volume}{429}},
  \bibinfo{pages}{277} (\bibinfo{year}{2004}).

\bibitem[{\citenamefont{J\"{o}rdens et~al.}(2008)\citenamefont{J\"{o}rdens,
  Strohmaier, G\"{u}nter, Moritz, and Esslinger}}]{JOR08}
\bibinfo{author}{\bibfnamefont{R.}~\bibnamefont{J\"{o}rdens}},
  \bibinfo{author}{\bibfnamefont{N.}~\bibnamefont{Strohmaier}},
  \bibinfo{author}{\bibfnamefont{K.}~\bibnamefont{G\"{u}nter}},
  \bibinfo{author}{\bibfnamefont{H.}~\bibnamefont{Moritz}}, \bibnamefont{and}
  \bibinfo{author}{\bibfnamefont{T.}~\bibnamefont{Esslinger}},
  \bibinfo{journal}{Nature} \textbf{\bibinfo{volume}{455}},
  \bibinfo{pages}{204} (\bibinfo{year}{2008}).

\bibitem[{\citenamefont{Schneider et~al.}(2008)\citenamefont{Schneider,
  Hackerm\"{u}ller, Will, Best, Bloch, Costi, Helmes, Rasch, and
  Rosch}}]{SCH08}
\bibinfo{author}{\bibfnamefont{U.}~\bibnamefont{Schneider}},
  \bibinfo{author}{\bibfnamefont{L.}~\bibnamefont{Hackerm\"{u}ller}},
  \bibinfo{author}{\bibfnamefont{S.}~\bibnamefont{Will}},
  \bibinfo{author}{\bibfnamefont{T.}~\bibnamefont{Best}},
  \bibinfo{author}{\bibfnamefont{I.}~\bibnamefont{Bloch}},
  \bibinfo{author}{\bibfnamefont{T.~A.} \bibnamefont{Costi}},
  \bibinfo{author}{\bibfnamefont{R.~W.} \bibnamefont{Helmes}},
  \bibinfo{author}{\bibfnamefont{D.}~\bibnamefont{Rasch}}, \bibnamefont{and}
  \bibinfo{author}{\bibfnamefont{A.}~\bibnamefont{Rosch}},
  \textbf{\bibinfo{volume}{322}}, \bibinfo{pages}{1520} (\bibinfo{year}{2008}).

\bibitem[{\citenamefont{Chin et~al.}(2006)\citenamefont{Chin, Miller, Liu,
  Stan, Setiawan, Sanner, Xu, and Ketterle}}]{CHI06}
\bibinfo{author}{\bibfnamefont{J.~K.} \bibnamefont{Chin}},
  \bibinfo{author}{\bibfnamefont{D.~E.} \bibnamefont{Miller}},
  \bibinfo{author}{\bibfnamefont{Y.}~\bibnamefont{Liu}},
  \bibinfo{author}{\bibfnamefont{C.}~\bibnamefont{Stan}},
  \bibinfo{author}{\bibfnamefont{W.}~\bibnamefont{Setiawan}},
  \bibinfo{author}{\bibfnamefont{C.}~\bibnamefont{Sanner}},
  \bibinfo{author}{\bibfnamefont{K.}~\bibnamefont{Xu}}, \bibnamefont{and}
  \bibinfo{author}{\bibfnamefont{W.}~\bibnamefont{Ketterle}},
  \bibinfo{journal}{Nature} \textbf{\bibinfo{volume}{443}},
  \bibinfo{pages}{961} (\bibinfo{year}{2006}).

\bibitem[{\citenamefont{Mandel et~al.}(2003)\citenamefont{Mandel, Greiner,
  Widera, Rom, H\"{a}nsch, and Bloch}}]{Mandel}
\bibinfo{author}{\bibfnamefont{O.}~\bibnamefont{Mandel}},
  \bibinfo{author}{\bibfnamefont{M.}~\bibnamefont{Greiner}},
  \bibinfo{author}{\bibfnamefont{A.}~\bibnamefont{Widera}},
  \bibinfo{author}{\bibfnamefont{T.}~\bibnamefont{Rom}},
  \bibinfo{author}{\bibfnamefont{T.~W.} \bibnamefont{H\"{a}nsch}},
  \bibnamefont{and} \bibinfo{author}{\bibfnamefont{I.}~\bibnamefont{Bloch}},
  \bibinfo{journal}{Nature} \textbf{\bibinfo{volume}{425}},
  \bibinfo{pages}{937} (\bibinfo{year}{2003}).

\bibitem[{\citenamefont{Est\'{e}ve et~al.}(2008)\citenamefont{Est\'{e}ve,
  Gross, Weller, Giovanazzi, and Oberthaler}}]{Esteve_Nature_2008}
\bibinfo{author}{\bibfnamefont{J.}~\bibnamefont{Est\'{e}ve}},
  \bibinfo{author}{\bibfnamefont{C.}~\bibnamefont{Gross}},
  \bibinfo{author}{\bibfnamefont{A.}~\bibnamefont{Weller}},
  \bibinfo{author}{\bibfnamefont{S.}~\bibnamefont{Giovanazzi}},
  \bibnamefont{and} \bibinfo{author}{\bibfnamefont{M.~K.}
  \bibnamefont{Oberthaler}}, \bibinfo{journal}{Nature}
  \textbf{\bibinfo{volume}{455}}, \bibinfo{pages}{1216} (\bibinfo{year}{2008}).

\bibitem[{\citenamefont{Weitenberg et~al.}(2011)\citenamefont{Weitenberg,
  Endres, Sherson, Cheneau, Schau\ss, Fukuhara, Bloch, and Kuhr}}]{singlespin}
\bibinfo{author}{\bibfnamefont{C.}~\bibnamefont{Weitenberg}},
  \bibinfo{author}{\bibfnamefont{M.}~\bibnamefont{Endres}},
  \bibinfo{author}{\bibfnamefont{J.~F.} \bibnamefont{Sherson}},
  \bibinfo{author}{\bibfnamefont{M.}~\bibnamefont{Cheneau}},
  \bibinfo{author}{\bibfnamefont{P.}~\bibnamefont{Schau\ss}},
  \bibinfo{author}{\bibfnamefont{T.}~\bibnamefont{Fukuhara}},
  \bibinfo{author}{\bibfnamefont{I.}~\bibnamefont{Bloch}}, \bibnamefont{and}
  \bibinfo{author}{\bibfnamefont{S.}~\bibnamefont{Kuhr}},
  \bibinfo{journal}{Nature} \textbf{\bibinfo{volume}{471}},
  \bibinfo{pages}{319} (\bibinfo{year}{2011}).

\bibitem[{\citenamefont{Stockman et~al.}(2002)\citenamefont{Stockman, Faleev,
  and Bergman}}]{PhysRevLett.88.067402}
\bibinfo{author}{\bibfnamefont{M.~I.} \bibnamefont{Stockman}},
  \bibinfo{author}{\bibfnamefont{S.~V.} \bibnamefont{Faleev}},
  \bibnamefont{and} \bibinfo{author}{\bibfnamefont{D.~J.}
  \bibnamefont{Bergman}}, \bibinfo{journal}{Phys. Rev. Lett.}
  \textbf{\bibinfo{volume}{88}}, \bibinfo{pages}{067402}
  (\bibinfo{year}{2002}).

\bibitem[{\citenamefont{Aeschlimann et~al.}(2010)\citenamefont{Aeschlimann,
  Bauer, Bayer, Brixner, Cunovic, Dimler, Fischer, Pfeiffer, Rohmer, Schneider
  et~al.}}]{Aeschlimann:ProceedingsOfTheNationalAcademyOfSciences:2010}
\bibinfo{author}{\bibfnamefont{M.}~\bibnamefont{Aeschlimann}},
  \bibinfo{author}{\bibfnamefont{M.}~\bibnamefont{Bauer}},
  \bibinfo{author}{\bibfnamefont{D.}~\bibnamefont{Bayer}},
  \bibinfo{author}{\bibfnamefont{T.}~\bibnamefont{Brixner}},
  \bibinfo{author}{\bibfnamefont{S.}~\bibnamefont{Cunovic}},
  \bibinfo{author}{\bibfnamefont{F.}~\bibnamefont{Dimler}},
  \bibinfo{author}{\bibfnamefont{A.}~\bibnamefont{Fischer}},
  \bibinfo{author}{\bibfnamefont{W.}~\bibnamefont{Pfeiffer}},
  \bibinfo{author}{\bibfnamefont{M.}~\bibnamefont{Rohmer}},
  \bibinfo{author}{\bibfnamefont{C.}~\bibnamefont{Schneider}},
  \bibnamefont{et~al.}, \bibinfo{journal}{Proceedings of the National Academy
  of Sciences} \textbf{\bibinfo{volume}{107}}, \bibinfo{pages}{5329}
  (\bibinfo{year}{2010}).

\bibitem[{\citenamefont{Sentenac and Chaumet}(2008)}]{PhysRevLett.101.013901}
\bibinfo{author}{\bibfnamefont{A.}~\bibnamefont{Sentenac}} \bibnamefont{and}
  \bibinfo{author}{\bibfnamefont{P.~C.} \bibnamefont{Chaumet}},
  \bibinfo{journal}{Phys. Rev. Lett.} \textbf{\bibinfo{volume}{101}},
  \bibinfo{pages}{013901} (\bibinfo{year}{2008}).

\bibitem[{\citenamefont{Kao et~al.}(2011)\citenamefont{Kao, Jenkins,
  Ruostekoski, and Zheludev}}]{PhysRevLett.106.085501}
\bibinfo{author}{\bibfnamefont{T.~S.} \bibnamefont{Kao}},
  \bibinfo{author}{\bibfnamefont{S.~D.} \bibnamefont{Jenkins}},
  \bibinfo{author}{\bibfnamefont{J.}~\bibnamefont{Ruostekoski}},
  \bibnamefont{and} \bibinfo{author}{\bibfnamefont{N.~I.}
  \bibnamefont{Zheludev}}, \bibinfo{journal}{Phys. Rev. Lett.}
  \textbf{\bibinfo{volume}{106}}, \bibinfo{pages}{085501}
  (\bibinfo{year}{2011}).

\bibitem[{\citenamefont{Lagendijk and Tiggelen}(1996)}]{Lagendijk1996143}
\bibinfo{author}{\bibfnamefont{A.}~\bibnamefont{Lagendijk}} \bibnamefont{and}
  \bibinfo{author}{\bibfnamefont{B.~A.~v.} \bibnamefont{Tiggelen}},
  \bibinfo{journal}{Physics Reports} \textbf{\bibinfo{volume}{270}},
  \bibinfo{pages}{143 } (\bibinfo{year}{1996}).

\bibitem[{\citenamefont{Morice et~al.}(1995)\citenamefont{Morice, Castin, and
  Dalibard}}]{MoriceEtAlPRA1995}
\bibinfo{author}{\bibfnamefont{O.}~\bibnamefont{Morice}},
  \bibinfo{author}{\bibfnamefont{Y.}~\bibnamefont{Castin}}, \bibnamefont{and}
  \bibinfo{author}{\bibfnamefont{J.}~\bibnamefont{Dalibard}},
  \bibinfo{journal}{Phys. Rev. A} \textbf{\bibinfo{volume}{51}},
  \bibinfo{pages}{3896} (\bibinfo{year}{1995}).

\bibitem[{\citenamefont{Ruostekoski and
  Javanainen}(1997{\natexlab{a}})}]{PhysRevA.55.513}
\bibinfo{author}{\bibfnamefont{J.}~\bibnamefont{Ruostekoski}} \bibnamefont{and}
  \bibinfo{author}{\bibfnamefont{J.}~\bibnamefont{Javanainen}},
  \bibinfo{journal}{Phys. Rev. A} \textbf{\bibinfo{volume}{55}},
  \bibinfo{pages}{513} (\bibinfo{year}{1997}{\natexlab{a}}).

\bibitem[{\citenamefont{Ruostekoski and
  Javanainen}(1997{\natexlab{b}})}]{PhysRevA.56.2056}
\bibinfo{author}{\bibfnamefont{J.}~\bibnamefont{Ruostekoski}} \bibnamefont{and}
  \bibinfo{author}{\bibfnamefont{J.}~\bibnamefont{Javanainen}},
  \bibinfo{journal}{Phys. Rev. A} \textbf{\bibinfo{volume}{56}},
  \bibinfo{pages}{2056} (\bibinfo{year}{1997}{\natexlab{b}}).

\bibitem[{\citenamefont{Van~Tiggelen}(1999)}]{vtiggelen99}
\bibinfo{author}{\bibfnamefont{B.}~\bibnamefont{van~Tiggelen}}, in
  \emph{\bibinfo{booktitle}{Diffuse Waves In Complex Media}}, edited by
  \bibinfo{editor}{\bibnamefont{{Fouque, JP}}}, \bibinfo{organization}{NATO,Sci
  Comm} (\bibinfo{publisher}{Springer}, \bibinfo{address}{
  Dordrecht, Netherlands}, \bibinfo{year}{1999}), vol. \bibinfo{volume}{531} of
  \emph{\bibinfo{series}{NATO Advanced Science Institutes Series, Series C,
  Mathematical And Physical Sciences}}, pp. \bibinfo{pages}{1--60}.

\bibitem{JAV03} J. Javanainen and J. Ruostekoski, \prl {\bf 91}, 150404 (2003).

\bibitem{MEK07} I. B. Mekhov, C. Maschler, and H. Ritsch, \prl {\bf 98}, 100402
(2007).

\bibitem[{\citenamefont{Ruostekoski et~al.}({2009})\citenamefont{Ruostekoski,
  Foot, and Deb}}]{ruo_light09}
\bibinfo{author}{\bibfnamefont{J.}~\bibnamefont{Ruostekoski}},
  \bibinfo{author}{\bibfnamefont{C.~J.} \bibnamefont{Foot}}, \bibnamefont{and}
  \bibinfo{author}{\bibfnamefont{A.~B.} \bibnamefont{Deb}},
  \bibinfo{journal}{{\prl}}
  \textbf{\bibinfo{volume}{{103}}}, \bibinfo{pages}{{170404}}
  (\bibinfo{year}{{2009}}).

\bibitem{rist_light} S. Rist, C. Menotti, and G. Morigi, \pra {\bf 81}, 013404 (2010).


\bibitem[{\citenamefont{Corcovilos et~al.}({2010})\citenamefont{Corcovilos,
  Baur, Hitchcock, Mueller, and Hulet}}]{hulet_light}
\bibinfo{author}{\bibfnamefont{T.~A.} \bibnamefont{Corcovilos}},
  \bibinfo{author}{\bibfnamefont{S.~K.} \bibnamefont{Baur}},
  \bibinfo{author}{\bibfnamefont{J.~M.} \bibnamefont{Hitchcock}},
  \bibinfo{author}{\bibfnamefont{E.~J.} \bibnamefont{Mueller}},
  \bibnamefont{and} \bibinfo{author}{\bibfnamefont{R.~G.} \bibnamefont{Hulet}},
  \bibinfo{journal}{{\pra}} \textbf{\bibinfo{volume}{{81}}},
  \bibinfo{pages}{{013415}} (\bibinfo{year}{{2010}}).

\bibitem[{\citenamefont{Douglas and Burnett}({2011})}]{bur_light2}
\bibinfo{author}{\bibfnamefont{J.~S.} \bibnamefont{Douglas}} \bibnamefont{and}
  \bibinfo{author}{\bibfnamefont{K.}~\bibnamefont{Burnett}},
  \bibinfo{journal}{{\pra}} \textbf{\bibinfo{volume}{{84}}},
  \bibinfo{pages}{{033637}} (\bibinfo{year}{{2011}}).

\bibitem[{\citenamefont{Douglas and Burnett}({2011})}]{burhall}
\bibinfo{author}{\bibfnamefont{J.~S.} \bibnamefont{Douglas}} \bibnamefont{and}
  \bibinfo{author}{\bibfnamefont{K.}~\bibnamefont{Burnett}},
  \bibinfo{journal}{{\pra}} \textbf{\bibinfo{volume}{{84}}},
  \bibinfo{pages}{{053608}} (\bibinfo{year}{{2011}}).

\bibitem{castinlattice} M. Antezza and Y. Castin, \prl {\bf 103}, 123903 (2009).

\bibitem[{\citenamefont{Morsch and Oberthaler}(2006)}]{Morsch06}
\bibinfo{author}{\bibfnamefont{O.}~\bibnamefont{Morsch}} \bibnamefont{and}
  \bibinfo{author}{\bibfnamefont{M.}~\bibnamefont{Oberthaler}},
  \bibinfo{journal}{Rev. Mod. Phys.} \textbf{\bibinfo{volume}{78}},
  \bibinfo{pages}{179} (\bibinfo{year}{2006}).

\bibitem[{\citenamefont{{Jackson}}(1998)}]{Jackson}
\bibinfo{author}{\bibfnamefont{J.~D.} \bibnamefont{{Jackson}}},
  \emph{\bibinfo{title}{Classical Electrodynamics}} (\bibinfo{publisher}{John
  Wiley \& Sons}, \bibinfo{address}{New York}, \bibinfo{year}{1998}).

\bibitem[{\citenamefont{Javanainen et~al.}(1999)\citenamefont{Javanainen,
  Ruostekoski, Vestergaard, and Francis}}]{JavanainenEtAlPRA1999}
\bibinfo{author}{\bibfnamefont{J.}~\bibnamefont{Javanainen}},
  \bibinfo{author}{\bibfnamefont{J.}~\bibnamefont{Ruostekoski}},
  \bibinfo{author}{\bibfnamefont{B.}~\bibnamefont{Vestergaard}},
  \bibnamefont{and} \bibinfo{author}{\bibfnamefont{M.~R.}
  \bibnamefont{Francis}}, \bibinfo{journal}{Phys. Rev. A}
  \textbf{\bibinfo{volume}{59}}, \bibinfo{pages}{649} (\bibinfo{year}{1999}).

\bibitem[{\citenamefont{Chomaz}(2011)}]{Chomaz11}
\bibinfo{author}{\bibfnamefont{L.}~\bibnamefont{Chomaz}},
\bibinfo{author}{\bibfnamefont{L.}~\bibnamefont{Corman}},
\bibinfo{author}{\bibfnamefont{T.}~\bibnamefont{Yefsah}},
\bibinfo{author}{\bibfnamefont{R.}~\bibnamefont{Desbuquois}},
 \bibnamefont{and}
  \bibinfo{author}{\bibfnamefont{J.}~\bibnamefont{Dalibard}},
  \bibinfo{journal}{arXiv:1112.3170} (\bibinfo{year}{2011}).

\bibitem[{\citenamefont{AlAssam}(2010)}]{AlAssam10}
\bibinfo{author}{\bibfnamefont{S.}~\bibnamefont{Al-Assam}},
  \bibinfo{author}{\bibfnamefont{R. A.}~\bibnamefont{Williams}},  \bibnamefont{and}
  \bibinfo{author}{\bibfnamefont{C.}~\bibnamefont{Foot}},
  \bibinfo{journal}{Phys. Rev. A} \textbf{\bibinfo{volume}{82}},
  \bibinfo{pages}{021604} (\bibinfo{year}{2010}).

  \bibitem[{\citenamefont{Shotter}(2008)}]{Shotter08}
\bibinfo{author}{\bibfnamefont{M.}~\bibnamefont{Shotter}},
\bibinfo{author}{\bibfnamefont{D.}~\bibnamefont{Trypogeorgos}},  \bibnamefont{and}
  \bibinfo{author}{\bibfnamefont{C.}~\bibnamefont{Foot}},
  \bibinfo{journal}{Phys. Rev. A} \textbf{\bibinfo{volume}{78}},
  \bibinfo{pages}{051602} (\bibinfo{year}{2008}).


\end{thebibliography}
\end{document}